\documentclass[10pt]{elsart}
\usepackage[dvips]{graphicx}
\renewcommand{\thefootnote}{\fnsymbol{footnote}}
\begin{document}
\begin{frontmatter}

\title{Possible background reductions in double beta decay experiments}

\author[IReS]{R.~Arnold},
\author[LAL]{C.~Augier},
\author[INEL]{J.~Baker},
\author[ITEP]{A.~Barabash$^{\;1)}$}\thanks{Corresponding author,
Institute of Theoretical and Experimental Physics, B.~Cheremushkinskaya 25,
117259 Moscow, Russia,  e-mail: barabash@vxitep.itep.ru,
tel.: 007 (095) 129-94-68, fax: 007 (095) 883-96-01},
\author[IReS]{O.~Bing},
\author[JINR]{V.~Brudanin},
\author[INEL]{A.J.~Caffrey},
\author[IReS]{E.~Caurier},
\author[LAL]{K.~Errahmane},
\author[LAL]{A.-I.~Etienvre}
\author[IReS]{J.L.~Guyonnet},
\author[CENBG]{F.~Hubert},
\author[CENBG]{Ph.~Hubert},
\author[CENBG]{C.~Jollet},
\author[LAL]{S.~Jullian},
\author[JINR]{O.~Kochetov},
\author[JINR]{V.~Kovalenko},
\author[LAL]{D.~Lalanne},
\author[CENBG]{F.~Leccia},
\author[LPC]{C.~Longuemare},
\author[IReS]{Ch.~Marquet},
\author[LPC]{F.~Mauger},
\author[MHC]{H.W.~Nicholson},
\author[SAGA]{H.~Ohsumi}
\author[CENBG]{F.~Piquemal},
\author[CFR]{J-L.~Reyss},
\author[LAL]{X.~Sarazin},
\author[JINR]{Yu.~Shitov},
\author[LAL]{L.~Simard},
\author[CTU]{I.~\v{S}tekl},
\author[JYV]{J.~Suhonen},
\author[MHC]{C.S.~Sutton},
\author[LAL]{G.~Szklarz},
\author[JINR]{V.~Timkin},
\author[JINR]{V.~Tretyak},
\author[ITEP]{V.~Umatov},
\author[CTU]{L.~V\'{a}la},
\author[ITEP]{I.~Vanyushin},
\author[ITEP]{V.~Vasilyev},
\author[CU]{V.~Vorobel},
\author[JINR]{Ts.~Vylov}

\address[CENBG]{CENBG, IN2P3-CNRS et Universit\'e de Bordeaux,
               33170 Gradignan, France}
\address[LPC]{LPC, IN2P3-CNRS et Universit\'e de Caen, 14032 Caen, France}
\address[JINR]{JINR, 141980 Dubna, Russia }
\address[CFR]{CFR, CNRS, 91190 Gif sur Yvette, France}
\address[INEL]{INEEL, Idaho Falls, ID 83415, U.S.A.}
\address[JYV]{JYV\"ASKYL\"A University, 40351 Jyv\"askyl\"a, Finland}
\address[ITEP]{ITEP, 117259  Moscow, Russia}
\address[LAL]{LAL, IN2P3-CNRS et Universit\'e Paris-Sud, 91405 Orsay, France}
\address[MHC]{MHC, South Hadley, Massachusetts 01075, U.S.A.}
\address[IReS]{IReS, IN2P3-CNRS et Universit\'e
              Louis Pasteur, 67037 Strasbourg, France.}
\address[CTU]{CTU FNSPE, Prague, 11519 Czech Republic.}
\address[CU]{Charles University, Prague, Czech Republic.}
\address[SAGA]{SAGA University, Saga, Saga 840-8502, Japan}
{\large NEMO Collaboration}

\date{ }

\begin{abstract}
The background induced by radioactive impurities of $^{208}\rm Tl$
and $^{214}\rm Bi$ in the source of the double beta experiment NEMO-3
has been investigated. New methods of
data analysis which decrease
the background from the  above mentioned contamination
 are identified.
The techniques can also be applied to other double beta decay experiments capable of
measuring independently the energies of the two electrons.

{\it PACS:} 23.40.-s, 14.80.Mz

\begin{keyword}
Double beta decay, neutrino, Molybdenum-100.
\end{keyword}
\end{abstract}
\end{frontmatter}

\renewcommand{\baselinestretch}{1.6}

\newpage

\section{ Introduction }
~
Neutrinoless double beta decay ($2\beta0\nu$)
is a problem of great interest
in particle physics (see \cite{Smi00,Bil01}).
Currently, there are
many experiments searching for this subject, and
new ones are
planned for the future (e.g. \cite{Eji01,Fio01}).
One of the most promising of the experiments is NEMO-3. It will study double beta
decay of different nuclei
with the primary objective to look for
the $2\beta0\nu$
decay of $^{100}\rm Mo$. The detector
is sensitive to decays with a half life on the order of
$\sim10^{25}$ years \cite{LAL94,TAO99}.

The greatest concern in double beta experiments is the background
which can mimic the double beta decay process.
The backgrounds can be classified by their origin. There
are external and internal backgrounds.

The external background
accounts for such things as cosmic rays,
$\gamma-$rays and neutrons coming from the walls
of the laboratory and parts of the detector.
To decrease this background, one
places the detector deep underground, applies passive
and/or active
shielding
and incorporates radiopure materials.
The data from the NEMO-2 detector and
Monte-Carlo (MC)
calculations predict that the external background in
the NEMO-3 detector for five years
of operation will be equivalent to zero \cite{TAO99,NEUTRON}.

The internal background is caused by the properties of the source material.
An unavoidable background originates from
two neutrino double beta decay ($2\beta2\nu$)
of the source ($^{100}\rm Mo$).
This kind of internal background cannot be completely separated from the $2\beta 0\nu$
decay mode due to the finite energy
resolution of the detector. Another source of internal background
is the radioactive impurities in the sample. The energy of the $\beta^{-}$ decay
of such nuclei as $^{214}\rm Bi$ and $^{208}\rm Tl$
is greater than the energy of the $2\beta$ transition
of $^{100}\rm Mo$ (3.03 MeV).
Therefore they produce especially
troublesome backgrounds.   The
background from $^{214}\rm Bi$ and $^{208}\rm Tl$
clearly depends on the purity of the sample. However, producing source
materials with high purity is a sophisticated
and expensive procedure. Often the desired level
of purity cannot be reached. This is why it is
useful to elaborate on methods of data
analysis which suppress the backgrounds of the radioactive impurities.

Farther along in this paper the particular mechanisms by
which $^{214}\rm Bi$  and $^{208}\rm Tl$
can produce events similar to $2\beta 0\nu$  decay are discussed.
The characteristics of each type of background are described and
a new analysis, which can suppress them, is
presented.

\section{ The NEMO-3 detector.}
\begin{figure*}
\begin{center}
\includegraphics[width=6.5cm]{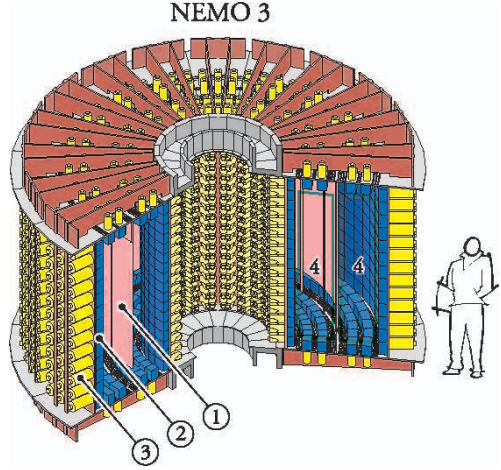}
\caption{The NEMO-3 detector. 1 - source foil; 2 - plastic
scintillator; 3 - low radioactive PMT; 4 - tracking chamber (6,180 octagonal
Geiger cells).
}
\label{nemo3}
\end{center}
\end{figure*}

~
The detector (Fig.~\ref{nemo3}) is cylindrical in design and it
is divided into
20 equal sectors. A thin (40-60 mg/cm$^2$) cylindrical
source foil is constructed from
either a metal film or powder, bound by an organic glue to mylar strips.
The source hangs between two concentric cylindrical tracking
volumes consisting of open octagonal drift cells operating in Geiger
mode. These cells run vertically and are staged in a 4, 2, and 3 row
pattern to optimise track reconstruction. The design of the 6,180 drift cells
calls for 50 $\mu$m  anode and cathode wires.

The
tracking volumes are surrounded
by a calorimeter made of 1,940 large blocks
of plastic scintillator coupled to very low
radioactivity 3" and 5" Hammamatsu photomultiplier tubes (PMTs).
At 1 MeV, the energy resolution,
which depends on the scintillator shape and the associated PMT, ranges
from 11\% to 14.5\% (FWHM), with a time resolution of 250 ps.

A solenoid producing a magnetic field ($\sim 30$ G)
surrounds the detector to reject ($e^+,e^-)$ pairs. External shielding
of 20 cm of low activity iron reduces gamma-ray fluxes and
thermal neutrons. A supplementary
shield (water tanks, wood  and polyethylene plates)
suppress the contribution of fast neutrons
to the external background.
At the depth of the experimental hall (4,800 m of water equivalent)
in the Modane Underground
Laboratory, the signal from cosmic rays is negligible.
Vigorous flushing of the air
in the hall reduces the radon levels to 10-20 Bq/m$^{3}$. The
presence of $^{214}$Bi decays in the detector from this level
of radon contamination is below that introduced by the PMTs.

\begin{figure*}
\begin{center}
\includegraphics[width=6.5cm]{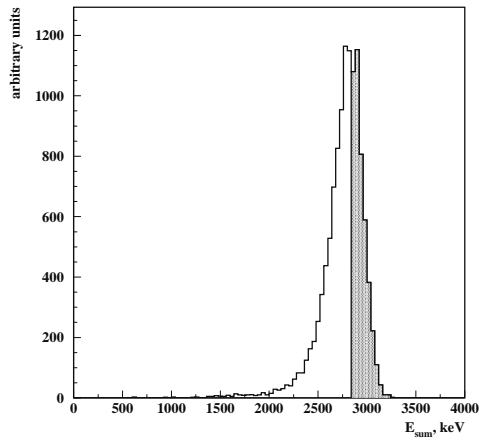}
\caption{Sum of the two electron energies (E$_{\rm sum}$) in
 $2\beta0\nu$ decay of $^{100} \rm Mo$.
High energy region
(2,840 to 3,240 keV) is highlighted.
}
\label{moesum}
\end{center}
\end{figure*}

An electron in the detector is defined by a track linking the
source foil and a fired scintillator.
A two-electron event (2e-event) is defined by two tracks which have
a common vertex in the
foil, and each track is associated with a fired scintillator.
The neutrinoless double beta decay gives rise to
two electrons which produce a 2e-event in the detector
with a line spectrum, because two electrons emitted carry all the
energy of the decay. However, due to energy
loss and the finite energy resolution of the calorimeter,
the summed electron spectrum of $2\beta0\nu$ events
has some width. The MC
calculations predict that the spectrum of $2\beta0\nu$ decays of
$^{100}\rm Mo$  in NEMO-3  will be as shown in Fig.~\ref{moesum}.

To suppress the background from $2\beta2\nu$ decay only
the 2e-events with energies within the window 2,840-3,240 keV are selected for further
analysis.  Hereafter, this window is referred to as
the high-energy region (HER).

A
 characteristic
 feature of the NEMO-3 detector
is that it can not only measure the whole
energy unleashed in the decay, but also the energy of each electron and the angle between
them.  A spectrum of the individual
electron energy in a 2e-event depends on its source
in the foil. This phenomenon and its effective analysis are
studied in the rest of the paper.

\section{ The primary sources of the internal backgrounds}
~

The internal background  estimations reported here, were made
with MC calculations. The calculations were
done with the help of the GEANT 3.21 package.

\begin{figure*}
\begin{center}
\includegraphics[width=6.5cm]{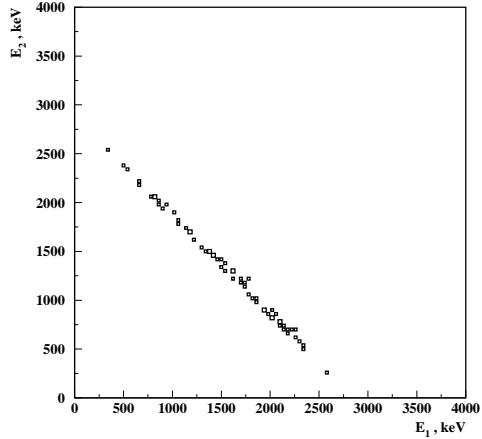}
\caption{2e-events in the HER
from $2\beta2\nu$ decay of $^{100} \rm Mo$
(295,000,000 $^{100} \rm Mo$ decays were simulated).
}
\label{mo2d2n}
\end{center}
\end{figure*}

The way by which the energy of 2e-event is distributed between the electrons
can be
well represented in a two-dimensional histogram (i.e. Fig.~\ref{mo2d2n}). 
   Along
the axis of these
histogram are the energies of electron 1 (E$_{1}$) and 2
(E$_{2}$). Thus, each 2e-event
looks like
a point.

The 2e-events from $2\beta2\nu$ decay of $^{100}\rm Mo$ in
the HER are shown in Fig.~\ref{mo2d2n}.
Notice that for the
majority of the events, the energy is divided almost
equally between the two electrons.

In the study of the background from the radioactive impurities in the source, it is
important to understand the exact mechanism by which the $\beta^{-}$ decay of $^{214}\rm Bi$
or $^{208}\rm Tl$  can be detected as a 2e-event
in the HER.  After scrutinising the results of
the MC calculations, three basic mechanisms were identified.

The first is $\beta^{-}$ decay accompanied by an electron
from the electron conversion 
(EC) mechanism.
In this case, two electrons with sufficient energy are emitted from a
common point.

The second process is M\"oller scattering
of the $\beta^{-}$ decay electron in the material
of the foil. The two electrons can be detected as
a 2e-event even if the $\beta^{-}$ decay
is followed by a transition of the daughter nucleus from an excited state to the
ground state (g.s.)
via a $\gamma$-ray emission. This occurs because the $\gamma$-ray
can escape the detector, given that the
detection efficiency of the scintillator is rather low (30-50$\%$).

The third possibility is that the second electron is produced via Compton
scattering of the $\gamma$-ray, which follows the $\beta^{-}$ decay. Again, as
in the previous case, this process can be interpreted as a 2e-event if the
scattered $\gamma$-ray is not detected.

\subsection{ The background from $^{214}\rm Bi$.}
\begin{figure*}
\begin{center}
\includegraphics[width=6.5cm]{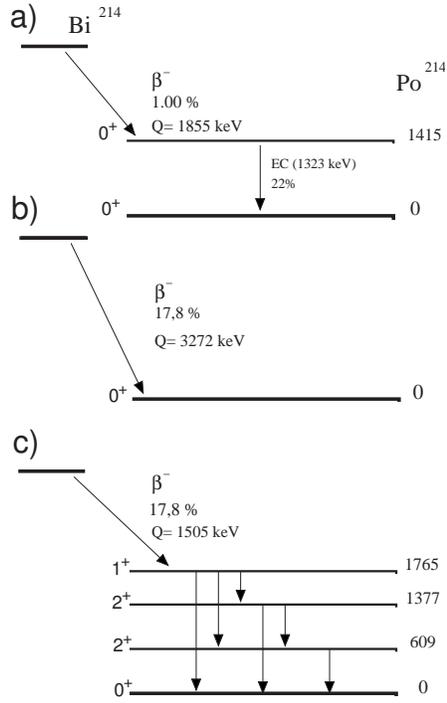}
\caption{$^{214}\rm Bi$ decay modes, which give the major contribution to the 2e-background in
the HER.
a) $\beta^{-}$ decay to the 1,415 keV level of $^{214}\rm Po$
and farther EC to the g.s. ($0.22\%$ of all decays);
b) $\beta^{-}$ decay to the g.s. of $^{214}\rm Po$ ($17.8\%$ of all decays);
c) $\beta^{-}$ decay to the 1,765 keV level of $^{214}\rm Po$
and its decay to the g.s.
($17.8\%$ of all decays).
}
\label{bidecay}
\end{center}
\end{figure*}

~
The $\beta^{-}$ decay of $^{214}\rm Bi$ produces 2e-events
via all of the above mentioned mechanisms.
These are the three dominate modes of the decay
which yield almost all the background from $^{214}\rm Bi$ in the HER.
The specific transitions are shown in Fig.~\ref{bidecay}.
The results of the background MC simulations are presented in Table 1.

\begin{figure*}
\begin{center}
\includegraphics[width=6.5cm]{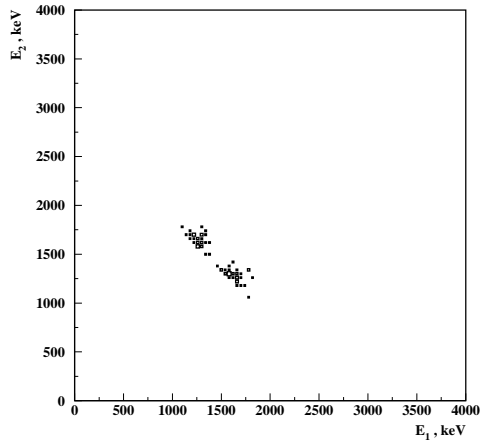}
\caption{2e-events in the HER
from $\beta^{-}$ decay of $^{214}\rm Bi$ to the
1,415 keV level of $^{214}\rm Po$ and then electron conversion to the g.s.
(Fig.~\ref{bidecay} a).
}
\label{biec2d}
\end{center}
\end{figure*}

The characteristic features of the background, produced by each of these processes, are as follows.
In the first process (Fig.~\ref{bidecay} a), a conversion electron with fixed energy (1,323 keV) is emitted.
One can see the conversion line in the
histogram of the corresponding 2e-events in Fig.~\ref{biec2d}.
Though the probability of this process is very low (0.22$\%$)
it contributes about half of the background from $^{214}\rm Bi$ in the HER (see Table 1).

\begin{figure*}
\begin{center}
\includegraphics[width=6.5cm]{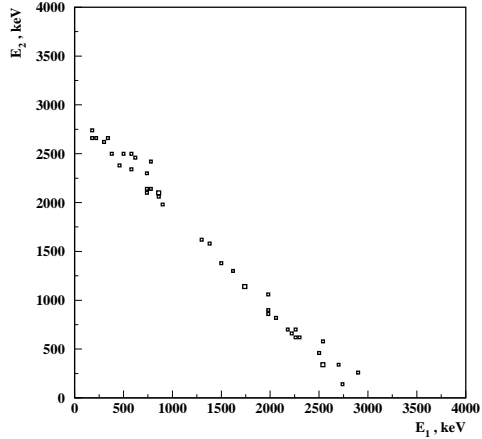}
\caption{2e-events in the HER from $\beta^{-}$ decay of $^{214}\rm Bi$ to the
g.s. of $^{214}\rm Po$
(Fig.~\ref{bidecay} b).
 }
\label{bi02d}
\end{center}
\end{figure*}

In the second process (Fig.~\ref{bidecay} b), the second electron is produced by M\"oller
scattering. Thus, there is a small momentum transfer and a small
scattering angle.  This can be observed in the 2e-events (Fig.~\ref{bi02d}).
One of the electrons in the pair has significantly higher energy than
the other.

\begin{figure*}
\begin{center}
\includegraphics[width=6.5cm]{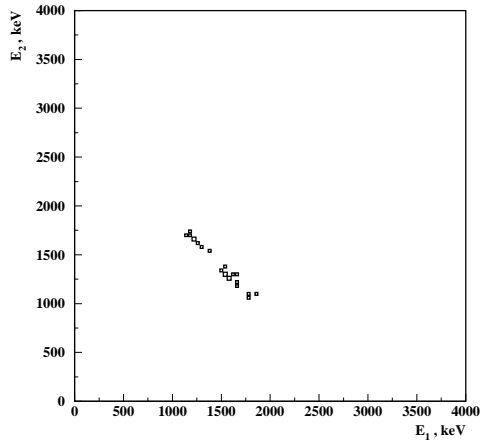}
\caption{2e-events in the HER
from $\beta^{-}$ decay of $^{214}\rm Bi$ to the
1,765 keV level of $^{214}\rm Po$ and electron conversion to the g.s.
(Fig.~\ref{bidecay} c).
}
\label{bi17652d}
\end{center}
\end{figure*}

Finally, in the third process (Fig.~\ref{bidecay} c),
the energies of the $\beta$-electron and the $\gamma$-ray are almost equal
($\sim1.5$ MeV).  Therefore the total energy of the two electrons is approximately 3 MeV
only if the second (Compton) electron acquires nearly all the energy of the $\gamma$-ray.
Thus, the corresponding 2e-events registered in the HER have two electrons with practically
the same energies (Fig.~\ref{bi17652d}).

\tabcolsep=0.05em
\begin{table}
\label{tbicontr}
\caption{
MC calculations of the background in the HER (2,840 to 3,240 keV) from
different $^{214}\rm Bi$ decay modes.}
\vspace{0.5cm}
\begin{center}
\begin{tabular}{|l|c|c|c|}
\hline
Decay mode  (See Fig.~\ref{bidecay})& Branching  &
Number of 2e-events$^{*}$&  Percentage of \\
~ &ratio&
~& $^{214}\rm Bi$ background \\

\hline
$\beta^{-}$(1,855 keV) plus EC(1,323 keV) & 0.0022 &$22.6\pm 2.7$& 47 \% \\
\hline
$\beta^{-}$(3,272 keV) & 0.178 &$12.7\pm 2.0$ & 26 \% \\
\hline
$\beta^{-}$(1,505 keV) plus $\gamma $ & 0.178 & $11\pm 2$& 22 \% \\
\hline
Sum &  &$46.3\pm 6.7$ & 96 \% \\
\hline
All decay modes  & 1.00 & $48\pm 7$ & 100 \% \\

\hline
\multicolumn{4}{l}{$^{*}$\rule{0pt}{11pt}\footnotesize Normalised to
14,450,000 $^{214}\rm Bi$ decays.}
\end{tabular}
\end{center}
\end{table}

\begin{figure*}
\begin{center}
\includegraphics[width=6.5cm]{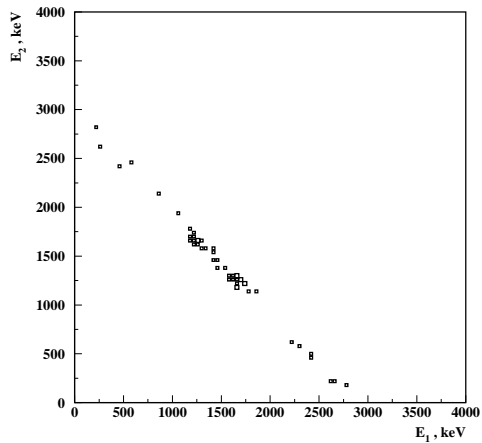}
\caption{ 2e-events in the HER
from $\beta^{-}$ decay of $^{214}\rm Bi$ for all decay modes.}
\label{bi2d}
\end{center}
\end{figure*}

A histogram of not only the three dominate decay modes but all
the background 2e-events from $^{214}\rm Bi$ is given in Fig.~\ref{bi2d}
for the HER.

\subsection{ The background from  $^{208}\rm Tl$ .}
\begin{figure*}
\begin{center}
\includegraphics[width=6.5cm]{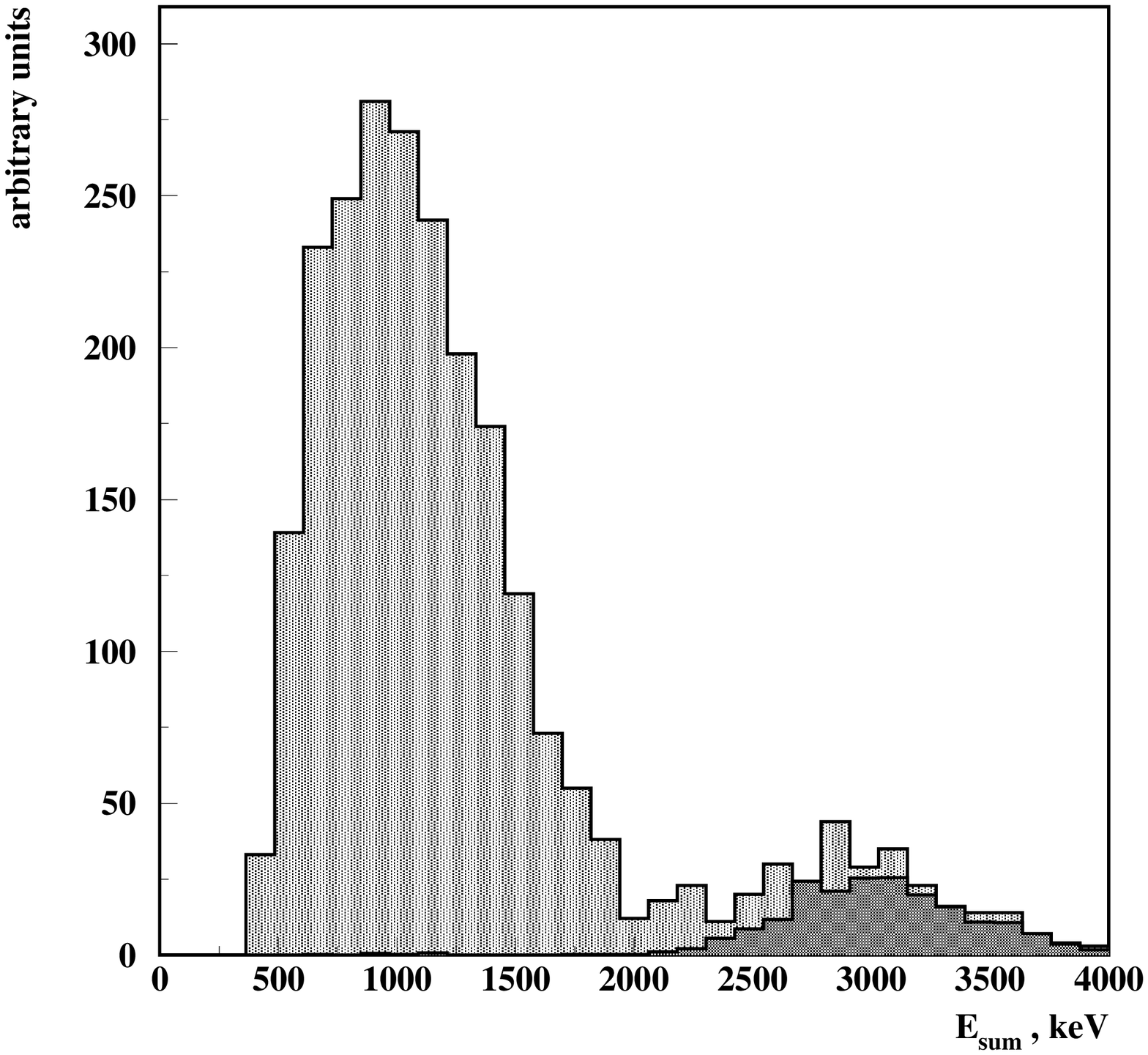}
\caption{The $\beta^{-}$ plus EC contribution  (darker area) to the 2e-background
from $^{208}\rm Tl$ (lighter area).
 }
\label{tl208ecc}
\end{center}
\end{figure*}

\begin{figure*}
\begin{center}
\includegraphics[width=6.5cm]{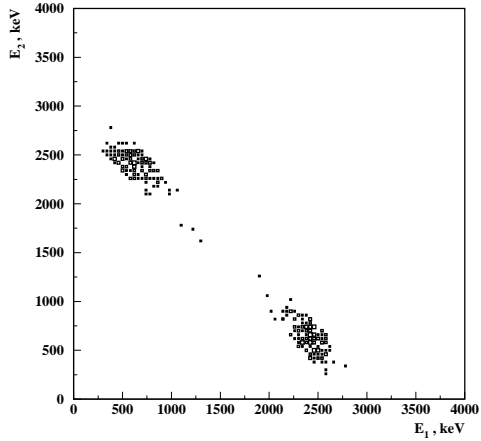}
\caption{ 2e-events in the HER
from $\beta^{-}$ decay of $^{208}\rm Tl$ to the
2,615 keV level of $^{208}\rm Pb$ and then EC to the g.s..
}
\label{tlec2d}
\end{center}
\end{figure*}

\begin{figure*}
\begin{center}
\includegraphics[width=6.5cm]{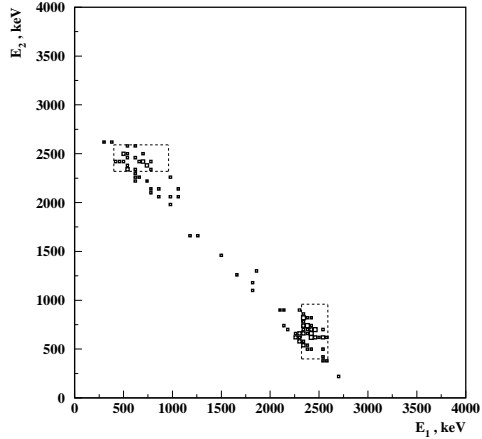}
\caption{2e-events in the HER
from $\beta^{-}$ decay of $^{208}\rm Tl$. The marked rectangles are
the proposed cut for background reduction (see text).
 }
\label{tl2082d}
\end{center}
\end{figure*}

~
The background of 2e-events in the HER from $^{208}\rm Tl$
comes mainly from the electron conversion mechanism. The
$\beta^{-}$ decay of $^{208}\rm Tl$ always occurs through
the excited state of a $^{208}\rm Pb$ nucleus with an energy of 2,615 keV.
This high energy and rather large probability of EC ($\sim 2.4\cdot10^{-3}$)
account for its large contribution
to the background in the HER. In Fig.~\ref{tl208ecc} one can see
2e-events from $^{208}\rm Tl$ and the contribution from the
$\beta^{-}$ decay process plus $\rm EC$.
In the HER it contributes up to $80\%$ of all $^{208}\rm Tl$ 2e-events.
As in the case with the background from $^{214}\rm Bi$, a conversion line can
be clearly seen in the
histogram of the 2e-events (Fig.~\ref{tlec2d},~\ref{tl2082d}).

\section{Results}
~

The MC results can be used to improve the
signal-to-noise ratio (S/N).
To do this, one should introduce new cuts, which take into account the difference in the energy
distributions among the electrons in the background 2e-events and
in the 2e-events from $2\beta0\nu$.

Currently, there are many
modes of $2\beta0\nu$ decay discussed in the literature \cite{Kla95,Sim98}.
Consider the two most favoured hypotheses: the existence of a non-zero
Majorana mass for the
neutrino and the admixture of the right hand current in the electro-weak
interaction. These two hypotheses
lead to different electron energy spectra, and thus they should be treated separately.

\subsection{$2\beta0\nu$ decay: Majorana neutrino mass}
~
The 2e-events from $2\beta0\nu$ decay, in the case of a non-zero
Majorana neutrino mass, are shown in Fig.~\ref{mo2d0n}. Their distribution is
similar
to
the one from $2\beta2\nu$ decay
(Fig.~\ref{mo2d2n}).
The 2e-events are symmetric from the point of view of
electrons energy.
Thus, one cannot distinguish $2\beta0\nu$ and $2\beta2\nu$ decays based
upon the energy of each electron in the pair.

The situation is different for the backgrounds from $^{214}\rm Bi$ and $^{208}\rm Tl$.
Since the major component of the background is caused by  the EC process, it can be suppressed
if one excludes the region where the conversion line peaks (Fig.~\ref{tl2082d}).
In Fig.~\ref{tl2082d} an example of the corresponding cut is given in the form
of rectangular areas to be excluded.
Table 2 presents the 
reduction
factors for various sizes of the excluded region for $^{208}\rm Tl$.

It is evident that one can get rid of a significant fraction of
the $^{208}\rm Tl$ background.
The 
reduction
of the $^{214}\rm Bi$ background is
less effective, because the conversion
line peaks in the region where the $2\beta0\nu$ decay is expected (Fig.~\ref{bi2d},\ref{mo2d0n})
and it also contributes less than half of the total $^{214}\rm Bi$ background.
The results for $^{214}\rm Bi$ are presented in Table 3.

\tabcolsep=10pt
\begin{table}
\label{ttlcut}
\caption{ MC data for the $^{208}\rm Tl$ background
reduction
 in the HER (Fig.~\ref{tl2082d}).}
\vspace{0.5cm}
\begin{center}
\begin{tabular}{|c|c|c|c|c|}
\hline
High energy & Low energy
 & $^{208}\rm Tl$ reduction    &$2\beta0\nu$  
reduction   &  S/N$^{*}$ \\
electron (keV)&
electron (keV) &  factor & factor &~ \\
\hline
no cut & no cut & 1.00& 1.00 &1.00 \\
\hline
2,320-2,590 & 400-960 & 0.37 & 0.90 & 2.43 \\
\hline
2,280-2,590 & 400-1,010 & 0.33 & 0.89 & 2.67 \\
\hline
2,230-2,590 & 430-900 & 0.26 & 0.85 & 3.31 \\
\hline
\multicolumn{5}{l}{$^{*}$\rule{0pt}{11pt}\footnotesize
Here S/N is the ratio between the $4^{th}$ and the $3^{rd}$ columns.}
\end{tabular}
\end{center}
\end{table}

\tabcolsep=0.1em
\begin{table}
\label{tbiec}
\caption{ MC data for the $^{214}\rm Bi$ background
reduction in the HER (Fig.~\ref{mo2d0n}).}
\vspace{0.5cm}
\begin{center}
\begin{tabular}{|c|c|c|c|c|c|}
\hline

High energy& Low energy
 & EC reduction &  $^{214}\rm Bi$
 reduction & $2\beta0\nu$ 
 reduction & S/N$^{*}$\\
electron (keV)&
electron (keV) &factor& factor &factor  & ~\\

\hline
no cut & no cut & 1.00 & 1.00 & 1.00 & 1.00  \\
\hline
1,320-1,720 & 1,170-1,310 & 0.45 & 0.56 & 0.90 & 1.60 \\
\hline
1,320-1,770 & 1,130-1,310 & 0.35 & 0.46 & 0.85 & 1.86 \\
\hline
1,320-1,770 & 1,090-1,320 & 0.25 & 0.42 & 0.81 & 1.94 \\
\hline
\multicolumn{5}{l}{$^{*}$\rule{0pt}{11pt}\footnotesize
Here S/N is the ratio between the $5^{th}$ and the $4^{th}$ columns.}
\end{tabular}
\end{center}
\end{table}

\subsection{$2\beta0\nu$ decay: right hand current}
\begin{figure*}
\begin{center}
\includegraphics[width=6.5cm]{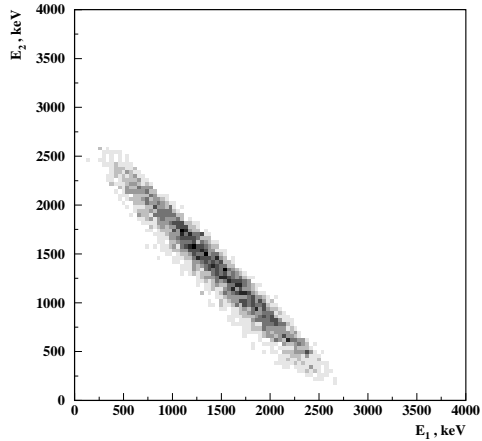}
\caption{2e-events from  $2\beta0\nu$ decay of $^{100} \rm Mo$
(Majorana neutrino mass).
}
\label{mo2d0n}
\end{center}
\end{figure*}

\begin{figure*}
\begin{center}
\includegraphics[width=6.5cm]{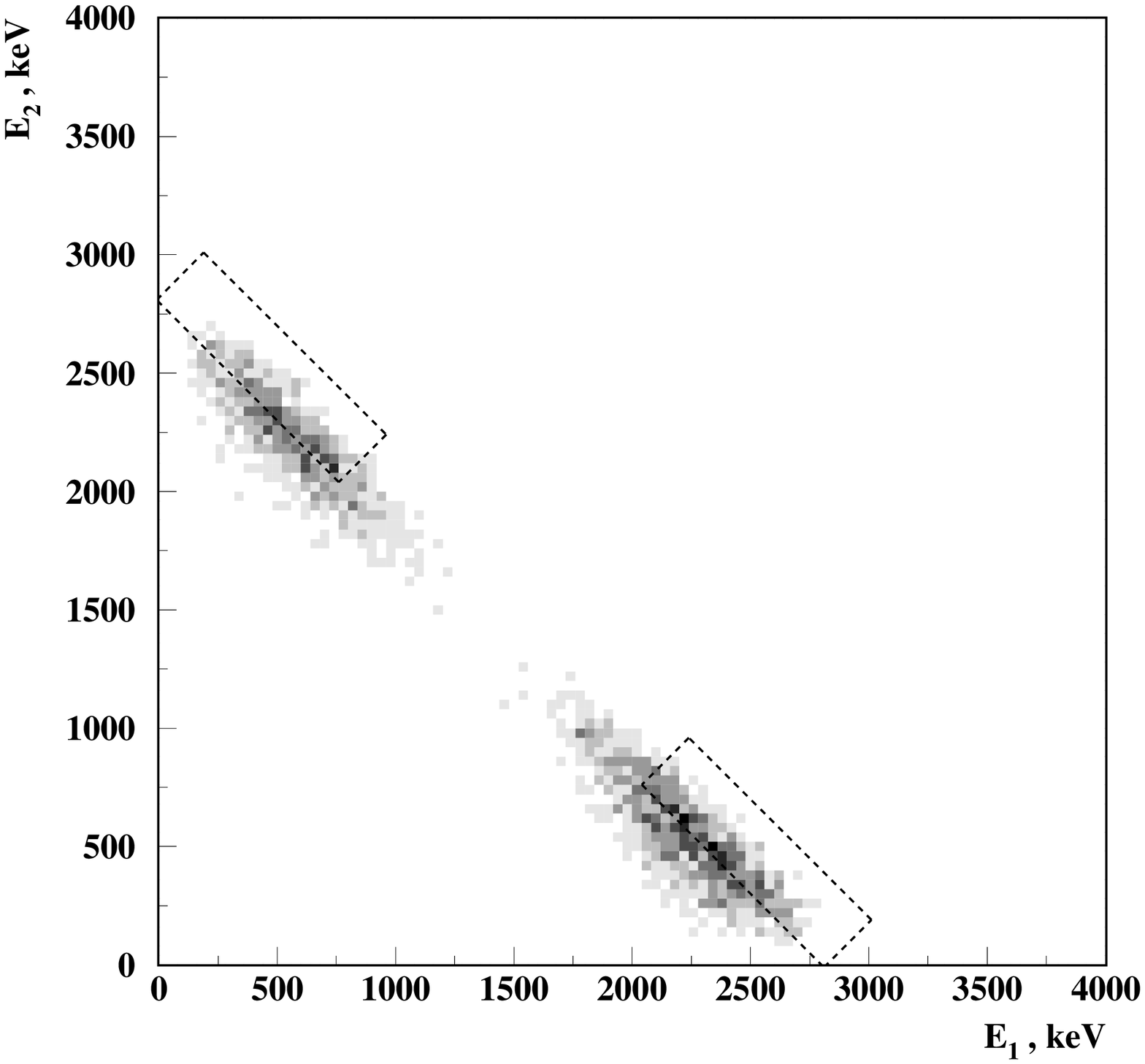}
\caption{2e-events from $2\beta0\nu$ decay of $^{100} \rm Mo$ 
(right hand current admixture mechanism).
The marked rectangles are
the proposed cut for background reduction (see text).
}
\label{mo2drc}
\end{center}
\end{figure*}

~
The distribution of 2e-events from $2\beta0\nu$ decay in the case of right hand
current admixture
in the electro-weak Lagrangian is shown in Fig.~\ref{mo2drc}. It
clearly differs from the case of the
Majorana neutrino
mass (Fig.~\ref{mo2d0n}).  Such a distribution allows
one to suppress efficiently the backgrounds
from $2\beta2\nu$ decay and $^{214}\rm Bi$, but makes it difficult
to suppress the $^{208}\rm Tl$
background. The suggested cut is to reject events, which do not lie in the rectangular
areas (Fig.~\ref{mo2drc}).
Table 4 presents these results.

\tabcolsep=10pt
\begin{table}
\label{mo0nrc}
\caption{MC data for the internal background reduction in $2\beta0\nu$ decay
with right hand current admixture (Fig.~\ref{mo2drc}).}
\vspace{0.5cm}
\begin{center}
\begin{tabular}{|c|c|c|c|c|}
\hline
Background&Decays & 2e-events &2e-events in cut region:& 
Reduction\\
~&simulated&in the HER&$\rm E_{sum}$ for 2,800 to 3,200&factor\\
~&~&~&$\mid\rm E_{1}-\rm E_{2}\mid$ for 1,280 to 2,820&~\\
\hline
$2\beta0\nu$&50,000&2330&2281&0.98\\
\hline
$2\beta2\nu$&295,000,000&63&43&0.68\\
\hline
$^{214}\rm Bi$&14,450,000&48&17&0.35\\
\hline
$^{208}\rm Tl$&2,170,000&105&96&0.91\\
\hline
\end{tabular}
\end{center}
\end{table}

\section{Conclusion}
~

In general one needs to gather as much information about
the event as possible to be able to suppress the backgrounds efficiently.
In double beta decay experiments
the measurement of the energies of each electron allows one to introduce
additional cuts,
which improve the S/N ratio. The final values of the cuts should
be tuned based on the properties of the detector and the background levels.

It was shown that the dominate part of the internal background comes from
the EC process associated with the radioactive
impurities ($^{214} \rm Bi$, $^{208}\rm Tl$) which can be 
efficiently suppressed.

The sensitivity of the
NEMO-3 detector to the right hand current admixture can be improved
by reducing the background from $^{214}\rm Bi$ and $2\beta2\nu$ decay.

The NEMO-3 detector can study
any double beta isotope that can be produced in foil form,
therefore similar analysis can be made on other isotopes and radioactive
contamination.

Though all the estimates were made for the NEMO-3 detector, qualitatively,
the results are valid for experiments where the energy of each electron
is measured. For example, this method can be applied to the data analysis
in existing (ELEGANT-V \cite{Eji98}, ITEP TPC \cite{ITE}) and future 
(MOON \cite{MOO},
DCBA \cite{DCB}) experiments.


\end{document}